\documentclass[reqno]{amsart}
\usepackage{hyperref}

\AtBeginDocument{{\noindent\small
\emph{Journal of Fractional Calculus and Applications},\\
Vol. 3(S). July,  11,   2012 (Proc. of the 4th. Symb. of Fractional Calculus and Applications)\\
No. 6, PP. 1 -- 8\newline
ISSN: 2090-5858.\\ http://www.fcaj.webs.com/
}
\vspace{9mm}}
\input{tcilatex}

\begin{document}
\title[\hfilneg JFCA-2012/4\hfil On the fractional-order games]{ A New
Technique for the Calculation of Effective Mesonic Potential at Finite
Temperature in the Logarithmic Quark-Sigma Model}
\author[M. Abu-Shady Herzallah\hfil JFCA-2012/4\hfilneg]{M. Abu-Shady}
\address{M. Abu-shady\\
Faculty of Science, Menofia University, Menofia, Egypt}
\email{abu\_shady\_1999@yahoo.com}
\address{ \\
}
\thanks{\textbf{Proc. of the 4th. Symb. of Frac. Calcu. Appl. Faculty of
Science Alexandria University, Alexandria, Egypt July, 11, 2012}}
\maketitle

\begin{abstract}
The logarithmic sigma model describes the interactions between quarks via
sigma and pion exchanges. The effective mesonic potential is extended to the
finite temperature and it is numerically calculated using n-midpoint rule.
Meson properties such as the phase transition, the sigma and pion masses,
and the critical point temperature are examined as functions of temperature.
The obtained results are compared with other approaches. We conclude that
the calculated effective potential is successfully to predict the meson
properties.
\end{abstract}

\section{Introduction}

The study of matter at very high temperature and densities is of interest
because of its relevance to particle physics and astrophysics. According to
the standard big bang model, it is believed that a series of phase
transitions happened at the early stages of the evolution of universe. The
QCD phase transition being one of them. The lattice QCD and effective field
theories are two main approaches to calculate the phase transition and meson
properties at finite temperature [1]. This subject has been under intense
theoretical study using various effective field theory models, such as the
Namu-Jona-Lasinio Model [2-4], a linear sigma model [5-8]. One of the
effective models in describing baryon properties is the linear sigma model,
which was suggested earlier by Gell-Mann and Levy [9] to describe the
nucleons interacting via sigma $(\sigma )$ and pion $(\mathbf{\pi })$
exchanges. At finite temperature, the model gives a good description of the
phase transition by using the Hartree approximation [1, 10-12] within the
Cornwall--Jackiw--Tomboulis (CJT) formalism [13]. However, there exists
serious difficulty concerning the renormalization of the CJT effective
action in the Hartree approximation. Baacke and Michalski [14] indicated
that the phase transition can be obtained beyond the large $N$ and Hartree
approximation through the systematic expansions are bases on the resummation
scheme by Cornwall-Jackiw-Tombonlis and 2PI scheme.

The aim of this work is to calculate the effective logarithmic mesonic
potential, the phase transition, and meson masses at finite temperature. The
logarithmic mesonic potential at zero temperature was suggested in Ref. [15]
to provide a good description of hadron properties. The\ used method is
different as a new technique in comparison with other works as in Refs.
[5-8]. In addition, the pressure is investigated as a function temperature.

This paper is organized as follows: In Sec. 2, the linear sigma model at
zero temperature and finite temperature are explained briefly. The numerical
calculations and discussion of the results are presented in Sec. 3. Summary
and conclusion is presented in Sec. 4.

\section{The Logarithmic Quark-Sigma Model}

\subsection{The Logarithmic Potential at zero Temperature}

The Lagrangian density of quark sigma model that describes the interactions
between quarks via\ the $\sigma -$and $\mathbf{\pi }-$meson exchange [15].
The Lagrangian density is, 
\begin{equation}
L\left( r\right) =i\overline{\Psi }\partial _{\mu }\gamma ^{\mu }\Psi +\frac{%
1}{2}\left( \partial _{\mu }\sigma \partial ^{\mu }\sigma +\partial _{\mu }%
\mathbf{\pi }.\partial ^{\mu }\mathbf{\pi }\right) +g\overline{\Psi }\left(
\sigma +i\gamma _{5}\mathbf{\tau }.\mathbf{\pi }\right) \Psi
-U^{T(0)}(\sigma ,\mathbf{\pi )},  \tag{1}
\end{equation}

with 
\begin{equation}
U^{T(0)}\left( \sigma ,\mathbf{\pi }\right) =\lambda _{1}^{2}(\sigma ^{2}+%
\mathbf{\pi }^{2})-\lambda _{2}^{2}\log (\sigma ^{2}+\mathbf{\pi }%
^{2})+m_{\pi }^{2}f_{\pi }\sigma ,  \tag{2}
\end{equation}%
$U^{T(0)}\left( \sigma ,\mathbf{\pi }\right) $ is the meson-meson
interaction potential where $\Psi ,\sigma $ and $\mathbf{\pi }$ are the
quark, sigma, and pion fields, respectively. In the mean-field approximation
the meson fields are treated as time-independent classical fields. This
means that we replace power and products of the meson fields by
corresponding powers and products of their expectation values. The
meson-meson interactions in Eq. (2) lead to hidden chiral $SU(2)\times SU(2)$
symmetry with $\sigma \left( r\right) $ taking on a vacuum expectation value
\ \ \ \ \ \ \ 
\begin{equation}
\ \ \ \ \ \ \left\langle \sigma \right\rangle =-f_{\pi },  \tag{3}
\end{equation}%
where $f_{\pi }=93$ MeV is the pion decay constant. The final \ term in Eq.
(2) is included to break the chiral symmetry. It leads to partial
conservation of axial-vector isospin current (PCAC). The parameters $\lambda
^{2},\nu ^{2}$ can be expressed in terms of $\ f_{\pi }$, the masses of
mesons as, 
\begin{equation}
\lambda _{1}^{2}=\frac{1}{4}(m_{\sigma }^{2}+m_{\pi }^{2}),  \tag{4}
\end{equation}%
\begin{equation}
\lambda _{2}^{2}=\frac{f_{\pi }^{2}}{4}(m_{\sigma }^{2}-m_{\pi }^{2}). 
\tag{5}
\end{equation}

\subsection{The Effective Mesonic Potential at Finite-Temperature\newline
}

In Eq. (2), The effective potential is extended to calculate the chiral
interacting mesons with quarks at finite temperature. The quarks are
considered a heat bath in local thermal equilibrium 
\begin{equation}
U_{eff}(\sigma ,\mathbf{\pi ,}T\mathbf{)}=U^{T(0)}(\sigma ,\mathbf{\pi )-}%
24T\dint \frac{d^{3}p}{\left( 2\pi \right) ^{3}}\ln (1+e^{\frac{-\sqrt{%
p^{2}+g^{2}(\sigma ^{2}+\mathbf{\pi }^{2})}}{T}}),  \tag{6}
\end{equation}%
\newline
the first term is the potential in the tree level defines in Eq. (2) and the
second term is for the chiral meson fields interacts with quarks at finite
temperature and zero-chemical potential. Non-zero values of the chiral
fields in the chiral broken phase dynamically generate a quark mass $%
m_{q}=gf_{\pi }$. The integration is taken over momentum volume (for
details, see Ref. [16]).

\section{Numerical Calculations and Discussion of Results}

\subsection{Numerical Calculations}

The purpose of this section is to calculate the effective mesonic potential $%
\sigma $ and $\mathbf{\pi }$- masses, and the pressure. \ We rewrite Eq. (6)
as follows%
\begin{equation}
U_{eff}(\sigma ,\mathbf{\pi ,}T\mathbf{)}=U^{T(0)}(\sigma ,\mathbf{\pi )-}%
\frac{12T}{\pi ^{2}}\int_{0}^{\infty }p^{2}dp(\ln (1+e^{\frac{-\sqrt{%
p^{2}+g^{2}(\sigma ^{2}+\mathbf{\pi }^{2})}}{T}}).  \tag{7}
\end{equation}%
Eq. (7) is written in the dimensionless form as follows%
\begin{equation}
U_{eff}(\sigma ,\mathbf{\pi ,}T\mathbf{)}=f_{\pi }^{4}[U^{T(0)}(\sigma
^{\prime },\mathbf{\pi }^{\prime }\mathbf{)-}\frac{12T^{\prime }}{\pi ^{2}}%
\int_{0}^{\infty }p^{^{\prime }2}(\ln e^{^{\left( -\frac{1}{T^{\prime }}%
\sqrt{P^{\prime 2}+g^{2}(\sigma ^{\prime 2}+\chi ^{\prime 2})}\right)
}}+1)dp^{\prime }],  \tag{8}
\end{equation}%
where%
\begin{equation}
U^{T(0)}(\sigma ^{\prime },\mathbf{\pi }^{\prime }\mathbf{)=}\lambda
_{1}^{^{\prime }2}(\sigma ^{\prime 2}+\mathbf{\pi }^{^{\prime }2})-\lambda
_{2}^{^{\prime }2}\log f_{\pi }^{2}(\sigma ^{^{\prime }2}+\mathbf{\pi }%
^{\prime 2})+m_{\pi }^{^{\prime }2}\sigma ^{\prime },  \tag{9}
\end{equation}%
and%
\begin{equation}
\lambda _{1}^{^{\prime }2}=\frac{1}{4}(m_{\sigma }^{^{\prime }2}+m_{\pi
}^{^{\prime }2}),\lambda _{2}^{\prime 2}=\frac{1}{4}(m_{\sigma }^{\prime
2}-m_{\pi }^{^{\prime }2}),  \tag{10}
\end{equation}%
where $\sigma ^{\prime },\mathbf{\pi }^{\prime },T^{\prime },m_{\pi
}^{^{\prime }},$and $m_{\sigma }^{^{\prime }}$ are in unit of $f_{\pi }.$
Therefore $U^{T(0)}(\sigma ^{\prime },\mathbf{\pi }^{\prime }\mathbf{)}$ is
the dimensionless form of $U^{T(0)}(\sigma ,\mathbf{\pi )}$. Substituting $%
p^{\prime }=-\ln y$ into Eq. (8), we obtain:

\begin{eqnarray*}
U_{eff}(\sigma ,\mathbf{\pi ,}T\mathbf{)} &=&f_{\pi }^{4}[U^{T(0)}(\sigma
^{\prime },\mathbf{\pi }^{\prime }\mathbf{)-} \\
&&-\frac{12T^{\prime }}{\pi ^{2}}\int_{0}^{1}\frac{(\ln y)^{2}}{y}\ln \left(
\exp \left( -\frac{1}{T^{\prime }}\sqrt{(\ln y)^{2}+g^{2}(\sigma ^{\prime 2}+%
\mathbf{\pi }^{\prime 2})}\right) +1\right) dy\allowbreak ],\ \ \ \ \ 
\end{eqnarray*}%
hence we can write the dimensionless form of $U_{eff}(\sigma ,\mathbf{\pi ,}T%
\mathbf{)}$ as follows:

\begin{eqnarray}
U_{eff}(\sigma ^{\prime },\mathbf{\pi }^{\prime }\mathbf{,}T^{\prime }%
\mathbf{)} &=&[U^{T(0)}(\sigma ^{\prime },\mathbf{\pi }^{\prime }\mathbf{)-}
\notag \\
&&-\frac{12T^{\prime }}{\pi ^{2}}\int_{0}^{1}\frac{(\ln y)^{2}}{y}\ln (%
\begin{array}{c}
\exp \left( -\frac{1}{T^{\prime }}\sqrt{(\ln y)^{2}+g^{2}(\sigma ^{\prime 2}+%
\mathbf{\pi }^{\prime 2})}\right) \\ 
+1)%
\end{array}%
dy\allowbreak ].\ \ \ \ \   \TCItag{11}
\end{eqnarray}%
By using midpoint rule, we obtain the approximate integral as follows:%
\begin{equation}
U_{eff}(\sigma ,\mathbf{\pi ,}T\mathbf{)}=f_{\pi }^{4}[U^{T(0)}(\sigma
^{\prime },\mathbf{\pi }^{\prime }\mathbf{)-}\frac{12T^{\prime }}{\pi ^{2}}%
A\ln \left( \exp \left( -\frac{1}{T^{\prime }}\sqrt{g^{2}\left( \sigma
^{\prime 2}+\mathbf{\pi }^{\prime 2}\right) +B}\right) +1\right) ],  \tag{12}
\end{equation}

where%
\begin{equation}
A=\frac{1}{n}\sum_{i=0}^{n}\frac{1}{\frac{1}{n}i+\frac{1}{2n}}\ln ^{2}\left( 
\frac{1}{n}i+\frac{1}{2n}\right) \allowbreak ,\ \ \ B=\sum_{i=0}^{n}\ln
^{2}\left( \frac{1}{n}i+\frac{1}{2n}\right) .  \tag{13}
\end{equation}%
(for details, see Refs. [17, 18]). In Ref. [19], the authors applied the
second derivation of the effective potential respect to $\sigma ^{\prime }$
and $\pi ^{\prime }$ to obtain the effective meson masses. Then, the first
derivative of\ the effective potential $U_{eff}(\sigma ^{\prime },\mathbf{%
\pi }^{\prime }\mathbf{,}T^{\prime }\mathbf{)}$ is given by%
\begin{equation}
\frac{\partial U_{eff}(\sigma ^{\prime },\mathbf{\pi }^{\prime }\mathbf{,}T%
\mathbf{)}}{\partial \sigma }=\frac{1}{f_{\pi }}\frac{\partial
U_{eff}(\sigma ^{\prime },\mathbf{\pi }^{\prime }\mathbf{,}T\mathbf{)}}{%
\partial \sigma ^{\prime }}=f_{\pi }^{3}[\frac{\partial U_{0}(\sigma
^{\prime },\mathbf{\pi }^{\prime }\mathbf{,}T\mathbf{)}}{\partial \sigma
^{\prime }}\mathbf{-}\frac{12T^{\prime }}{\pi ^{2}}\frac{d_{1}d_{2}}{%
d_{3}+d_{3}d_{2}}],  \tag{14}
\end{equation}%
where%
\begin{eqnarray}
d_{1} &=&-Ag^{2}\sigma ^{\prime },\ \ d_{2}=\exp \left( -\frac{1}{T^{\prime }%
}\sqrt{B+g^{2}(\sigma ^{\prime 2}+\mathbf{\pi }^{\prime 2})}\right) ,\ \  
\notag \\
d_{3}\ &=&T^{\prime }\sqrt{B+g^{2}(\sigma ^{\prime 2}+\mathbf{\pi }^{\prime
2})}.  \TCItag{15}
\end{eqnarray}%
Then, we obtain the effective sigma mass as follows%
\begin{equation}
m_{\sigma }^{2}(T)=\frac{\partial ^{2}U_{eff}(\sigma ^{\prime },\mathbf{\pi }%
^{\prime }\mathbf{,}T\mathbf{)}}{f_{\pi }^{2}\partial \sigma ^{\prime 2}}%
=f_{\pi }^{2}[\frac{\partial ^{2}U_{0}(\sigma ^{\prime },\mathbf{\pi }%
^{\prime }\mathbf{,}T\mathbf{)}}{\partial \sigma ^{\prime 2}}\mathbf{-}\frac{%
12T}{\pi ^{2}}^{\prime }\frac{\partial }{\partial \sigma ^{\prime }}(\frac{%
d_{1}d_{2}}{d_{3}+d_{3}d_{2}})],  \tag{16}
\end{equation}%
\begin{equation}
m_{\sigma }(T)=f_{\pi }[\frac{\partial ^{2}U_{0}(\sigma ^{\prime },\mathbf{%
\pi }^{\prime }\mathbf{,}T\mathbf{)}}{\partial \sigma ^{\prime 2}}\mathbf{-}%
\frac{12T^{\prime }}{\pi ^{2}}\frac{\partial }{\partial \sigma ^{\prime }}(%
\frac{d_{1}d_{2}}{d_{3}+d_{3}d_{2}})]^{\frac{1}{2}},  \tag{17}
\end{equation}%
where%
\begin{eqnarray}
\frac{\partial }{\partial \sigma ^{\prime }}(\frac{d_{1}d_{2}}{%
d_{3}+d_{3}d_{2}}) &=&d_{1}\frac{d_{2}^{\prime }}{d_{3}+d_{2}d_{3}}+d_{2}%
\frac{d_{1}^{\prime }}{d_{3}+d_{2}d_{3}}-  \notag \\
&&d_{1}\frac{d_{2}}{\left( d_{3}+d_{2}d_{3}\right) ^{2}}\allowbreak \left(
d_{3}^{\prime }+d_{2}d_{3}^{\prime }+d_{3}d_{2}^{\prime }\right) , 
\TCItag{18}
\end{eqnarray}%
with%
\begin{equation}
d_{1}^{\prime }=\frac{\partial d_{1}}{\partial \sigma ^{\prime }}%
=\allowbreak -Ag^{2},  \tag{19}
\end{equation}

\begin{equation}
d_{2}^{\prime }=\frac{\partial d_{2}}{\partial \sigma ^{\prime }}%
=\allowbreak -\frac{1}{T^{\prime }}g^{2}\frac{\sigma ^{\prime }}{\sqrt{%
B+g^{2}(\sigma ^{\prime 2}+\mathbf{\pi }^{\prime 2})}}\exp \left( -\frac{1}{T%
}\sqrt{B+g^{2}(\sigma ^{\prime 2}+\mathbf{\pi }^{\prime 2})}\right)
\allowbreak ,  \tag{20}
\end{equation}%
\begin{equation}
d_{3}^{\prime }=\frac{\partial d_{3}}{\partial \sigma ^{\prime }}%
=\allowbreak Tg^{2}\frac{\sigma ^{\prime }}{\sqrt{B+g^{2}(\sigma ^{\prime 2}+%
\mathbf{\pi }^{\prime 2})}}.  \tag{21}
\end{equation}%
Similarly, we obtain the effective pion mass as follows%
\begin{equation}
m_{\pi }(T)=f_{\pi }[\frac{\partial ^{2}U(\sigma ^{\prime },\mathbf{\pi }%
^{\prime })}{\partial \mathbf{\pi }^{\prime 2}}-.\frac{12}{\pi ^{2}}%
T^{\prime }\frac{\partial }{\partial \mathbf{\pi }^{\prime }}(\frac{%
d_{4}d_{2}}{d_{3}+d_{3}d_{2}})]^{\frac{1}{2}},  \tag{22}
\end{equation}%
where 
\begin{equation}
d_{4}=-Ag^{2}\mathbf{\pi }^{\prime },  \tag{23}
\end{equation}

\begin{eqnarray}
\frac{\partial }{\partial \mathbf{\pi }^{\prime }}(\frac{d_{4}d_{2}}{%
d_{3}+d_{3}d_{2}}) &=&d_{2}\frac{d_{4}^{\prime }}{d_{3}+d_{2}d_{3}}+d_{4}%
\frac{d_{2}^{\prime }}{d_{3}+d_{2}d_{3}}-  \notag \\
&&d_{2}\frac{d_{4}}{\left( d_{3}+d_{2}d_{3}\right) ^{2}}\allowbreak \left(
d_{3}^{\prime }+d_{2}d_{3}^{\prime }+d_{3}d_{2}^{\prime }\right) , 
\TCItag{24}
\end{eqnarray}%
where%
\begin{equation}
d_{4}^{\prime }=\frac{\partial d_{4}}{\partial \mathbf{\pi }^{\prime }}%
=\allowbreak -Ag^{2},  \tag{25}
\end{equation}%
$\ \ $%
\begin{equation}
d_{2}^{\prime }=\frac{\partial d_{2}}{\partial \mathbf{\pi }^{\prime }}=-%
\frac{1}{T}g^{2}\frac{\mathbf{\pi }^{\prime }}{\sqrt{B+g^{2}(\sigma ^{\prime
2}+\mathbf{\pi }^{\prime 2})}}\exp \left( -\frac{1}{T}\sqrt{B+g^{2}(\sigma
^{\prime 2}+\mathbf{\pi }^{\prime 2})}\right) ,  \tag{26}
\end{equation}%
$\allowbreak $%
\begin{equation}
d_{3}^{\prime }=\frac{\partial d_{3}}{\partial \mathbf{\pi }^{\prime }}%
=\allowbreak Tg^{2}\frac{\mathbf{\pi }^{\prime }}{\sqrt{B+g^{2}(\sigma
^{\prime 2}+\mathbf{\pi }^{\prime 2})}}.  \tag{27}
\end{equation}%
In Ref. [16], the pressure is given in the dimensionless form as follow:%
\begin{equation}
P^{\prime }(\sigma ^{\prime },\mathbf{\pi }^{\prime },T)=U^{T(0)}(\sigma
^{\prime },\mathbf{\pi }^{\prime }\mathbf{)}-U_{eff}(\sigma ^{\prime },%
\mathbf{\pi }^{\prime }\mathbf{,T}^{\prime }\mathbf{),}  \tag{28}
\end{equation}

\section{$\allowbreak $Results and Discussion}

In Eq. (7), The integration is solved by the n-midpoint algorithm and the
index n is taken $n=1000$ to get a good accuracy for a numerically
integration. In Ref. [20], the authors used a different method for
calculating the integration in the effective potential and they obtained it
as a series of $M^{2}=m^{2}+\frac{\lambda }{2}\phi ^{2}$ ,\ where the $m,$ $%
\lambda $ are parameters of the model. The difficulty in this potential is
to determine a critical temperature $T_{c}.$ The two terms were taken only
in the expression of the potential since the increase of terms more\ than
two terms will be the critical point is a complex value, leading $T_{c}$ is
not a physics quantity. In Refs. [1-4], authors used double bubble graphs
instead of summing infinite set of daisy and superdaisy graphs using the
tree level propagators. Therefore, the Feynman diagrams are needed. In the
present work, The n-midpiont rule is used to avoid the difficulties in the
above approaches. Hence, we did not need to apply Feynman diagrams. The
parameters of the model such as $m_{\pi }=140$ MeV$,$ $m_{\sigma }=600$ MeV$%
,~f_{\pi }=93$ MeV, and coupling constant $g$ \ at zero-temperature are used
as the initial parameters at the finite temperature. In addition, the
effective pion and sigma masses are obtained as a second derivative respect
to meson field. This method is used extensively in other works such as in
the Ref. [19]. \ In this section, we examine the meson properties and the
phase transition which depend on the calculation of the effective mesonic
potential at finite temperature. We replaced the normal potential in the
linear sigma model [5] at zero temperature by logarithmic mesonic potential
[15]. In our previous work [15]. The logarithmic potential was successfully
to predict hadron properties at zero temperature.

In Figs. (1, 2, 3), we investigate the behavior of the sigma and pion masses
as functions of temperature. Moreover, the effect of the sigma mass and the
coupling constant $g$ on a critical point temperature in the presence of the
explicit symmetry breaking ($m_{\pi }\neq 0)$ and the chiral limit( $m_{\pi
}=0$) is investigated. In Fig. 1, the sigma and pion masses are plotted as
functions of temperature at the presence of the explicit symmetry breaking
term. The sigma mass decreases with increasing temperature and the pion mass
increases with increasing temperature. The two curves crossed at a critical
point temperature $T_{c}$ where the sigma and pion masses have the same
massive value. At $m_{\sigma }=600$ MeV$,$ we find the critical point
temperature $T_{c}$ =233 MeV. By increasing the sigma mass up to $m_{\sigma
}=700$ MeV leads to $T_{c}$ = 309 MeV. In comparison with lattice QCD
results [19], the critical point temperature is found in the range $%
T_{c}=100\ $to 300 MeV. Therefore the present result is in agreement with
lattice QCD results. Nemoto et al. [19] calculated the critical point
temperature equal 230 MeV using the original sigma model. Abu-shady [5]
calculated the critical point temperature equal 226 MeV in the original
sigma model. Therefore, the present result is in good agreement with Refs.
[5, 19]. In Fig. 2, the sigma and pion masses are plotted as functions of
temperature in the chiral limit ( $m_{\pi }=0$). A similar behavior is
obtained as in the case of the explicit symmetry term expect the pion mass
equal zero at zero temperature. In comparison with original sigma model as
in Refs. [1, 5, 19], we found that the behavior of the sigma and pion masses
are in good agreement with Refs. [1, 5, 19]. In Fig. 3, we examine the
effect of coupling constant $g$ on the critical point temperature. We find
that an increase on the coupling constant $g$ leads to decrease in the
critical point temperature g = 3.76 to g = 4.48 corresponding to $T_{c}=$
243 MeV to $T_{c}=191$ MeV$,$ respectively. In Fig. (4), the effective
potential is plotted as a function of temperature. We note that the
potential decreases with increasing temperature. In order to get more
insight into the nature of the phase transition and verify that the order
phase transition is a second-order phase transition. We calculate the
effective potential $U_{eff}(\Phi \mathbf{,}T\mathbf{)}$ as a function of
phase transition ($\Phi =\sqrt{\sigma ^{2}+\mathbf{\pi }^{2}})$ as seen in
Fig. (4). The shape of the potential confirms that\ the phase transition is
the second-order. Since it exhibits can degenerate one minima at $\Phi \neq 0
$. The indication of the second-order phase transition has been reported in
many works [14, 21- 23]. Also, We note the strong increase in the
temperature T which unchanged the shape of the potential

Next, we need to examine the effect of finite temperature on the behavior of
pressure where the quarks takes as a heat bath. In Fig. (5), the pressure is
plotted as a function of temperature. \ The pressure increases with
increasing temperature. Also, we note that the pressure value at
lower-values of temperature is not sensitive in comparison with the pressure
value at the higher-values of temperature. Hence the effect of largest
values of temperature is more affected on the value of pressure. Berger and
Christov [24] found that the pressure increases with increasing temperature
in hot medium using the NJL model in the mean field approximation. The
present behavior is in agreement with Ref. [24].

\section{ Summary and Conclusion}

In this work, the effective mesonic potential is calculated by using
n-midpoint algorithm in the logarithmic quark model. The meson properties,
the phase transition, and the pressure are calculated using the effective
mesonic potential. We summarized the following points:The behavior of sigma
and pion masses as functions of temperature are investigated. A comparison
with original sigma model is presented. The increase of sigma mass ($%
m_{\sigma })~$leads to increase the critical point temperature. The increase
of the coupling constant ($g)$ leads to decrease the critical point
temperature. The phase transition is predicted as a second-order phase
transition which agrees with other works. The critical point temperature and
the pressure are calculated and are in agreement with other works.
Therefore, the calculated effective logarithmic potential is successful to
predict the meson masses, the phase transition, and the pressure at finite
temperature.

\end{document}